\documentclass[aps,prl,reprint,superscriptaddress,longbibliography]{revtex4-2}

\usepackage{amsmath,amssymb,bm}
\usepackage{graphicx}
\usepackage{xcolor}
\usepackage[colorlinks=true,citecolor=blue!55!black,
            linkcolor=blue!55!black,urlcolor=blue!55!black]{hyperref}

\usepackage{booktabs}

\newcommand{\PG}{P_{\mathrm G}}
\newcommand{\Id}{\mathbb I}
\newcommand{\ket}[1]{|#1\rangle}
\newcommand{\bra}[1]{\langle#1|}

\begin{document}

\title{Projecting Fermions into Exact Scar Towers:\\ 
Local Protection across Phase Transitions}

\author{Wenjie Wei}
\affiliation{Beijing National Laboratory for Condensed Matter Physics and Institute of Physics,
Chinese Academy of Sciences, Beijing 100190, China}

\author{Kun Jiang}
\email{jiangkun@iphy.ac.cn}
\affiliation{Beijing National Laboratory for Condensed Matter Physics and Institute of Physics,
Chinese Academy of Sciences, Beijing 100190, China}
\affiliation{School of Physical Sciences, University of Chinese Academy of Sciences,
Beijing 100190, China}

\author{Hui-Ke Jin}
\email{jinhk@shanghaitech.edu.cn}
\affiliation{School of Physical Science and Technology,
ShanghaiTech University, Shanghai 201210, China}

\date{\today}

\begin{abstract}
We construct exact scar towers in $SO(n)$ ($n\ge3$) spin chains by Gutzwiller projection of $n$-flavor flat-band Kitaev chains. The projected roots have only singlet and antisymmetric components on each bond. These local constraints allow a staggered flavor flip to generate an equidistant tower.  The root and its first descendant exhaust the two-site configurations that are visited by the tower, leaving an orthogonal bond subspace unvisited. Deformations within this subspace can break $SO(n)$ symmetry and
reshape the low-energy physics without altering the exact scar tower. A polarization-selective deformation drives an exact first-order transition from the symmetry-protected topological phase to a ferromagnetic phase. The tower retains its wave functions and absolute energies across the phase transition, including at the gapless point. Exact scars therefore need not inherit the symmetries, order parameters, or criticality of the ground state.
\end{abstract}

\maketitle

\textit{Introduction.}
Quantum many-body scars (QMBS) are exceptional nonthermal eigenstates embedded in otherwise thermalizing spectra that violate the eigenstate thermalization hypothesis (ETH)~\cite{Deutsch1991,Srednicki1994,Rigol2008,Serbyn2021,Moudgalya_2022,Chandran2023}. First highlighted by persistent quench oscillations in Rydberg-atom arrays~\cite{Bernien2017} and the kinetically constrained PXP model~\cite{Turner2018,Turner2018_prb,Ho2019,Khemani2019,Lin2019,Choi2019}, scarred dynamics have since been placed on an exact footing through equidistant eigenstate towers in the Affleck--Kennedy--Lieb--Tasaki (AKLT) chain~\cite{Affleck1987,Moudgalya2018,Moudgalya2018_2}, spin-1 $XY$ magnets~\cite{Schecter2019}, and related models~\cite{Mark2020,Mark2020_2,Moudgalya2020_eta,ODea2020,Pakrouski2020,Ren2021}. As QMBS have evolved from a phenomenological curiosity
into a general paradigm for weak ergodicity breaking,
attention has shifted toward identifying the underlying
protective mechanisms, including projector-embedding
constructions~\cite{Shiraishi2017,Omiya2023},
quasisymmetry-based approaches~\cite{Ren2021},
commutant-algebra methods~\cite{Moudgalya2024},
and MPS-based frameworks~\cite{Moudgalya2020_mps},
which enable scar states to coexist with thermalizing spectra.

Intriguingly, although these scar towers govern coherent nonthermal dynamics at finite energy density, their exact constructions are typically built on top of specific symmetric or topological ground states. This raises a fundamental question: how are high-energy quantum scars related to the conventional hallmarks of quantum phases, such as symmetries, order parameters, and low-energy criticality~\cite{Yao2022,Daniel2023}? 
Systematically addressing this question requires isolating exact scar towers from the interactions that govern the low-energy phase diagram. Previous studies have shown that local terms confined to an unvisited subspace can break global symmetries and modify the bulk spectrum without affecting the scar tower~\cite{Moudgalya2020_mps,Chattopadhyay2020,Mark2020,ODea2020,Pakrouski2026}. Nevertheless, such constructions are usually tied to the algebraic properties of target spin wave functions, and their robustness across changes of aforementioned hallmarks remains largely unexplored. This leads to a natural question: can microscopic kinetic constraints completely decouple an exact scar tower from the low-energy phase structure?

\begin{figure*}[!t]
 \centering
 \includegraphics[width=\textwidth]{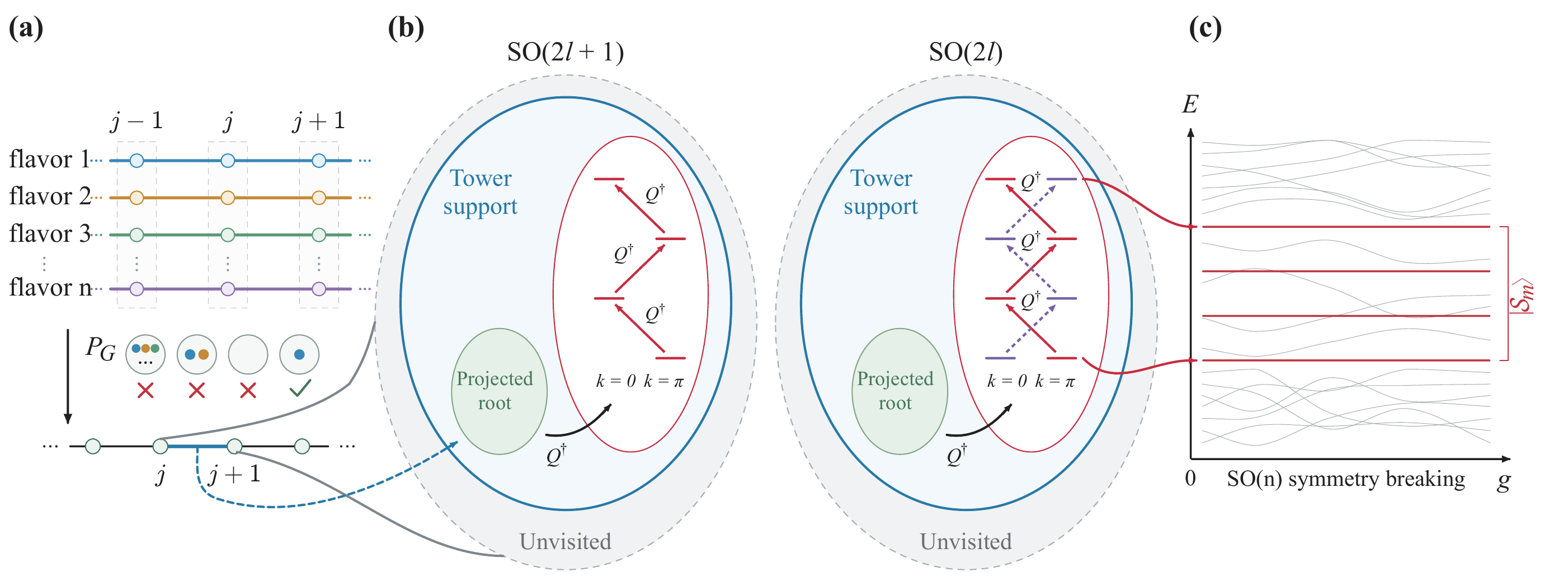}
 \caption{
 Local design of exact scar Hamiltonians from Gutzwiller-projected partons.
 (a) Single-occupancy projection of $n$ flat-band Kitaev chains. The Gutzwiller operator $\PG$ eliminates multiply occupied configurations, enforcing a one-parton-per-site constraint that maps Majorana-dimer pairing to physical spin degrees of freedom.
 (b) Tower generation. Roots (green) are mapped to scarred towers (red/purple) by the staggered ladder operator $Q^\dagger$. Magnifying a two-site bond reveals that the root and first descendant exhaust the tower's local support (blue boundary), isolating an unvisited orthogonal subspace (gray).
 (c) Deformations. Local deformation terms characterized by a strength $g$ acting only within the unvisited subspace reshape the thermal background (gray curves) while leaving the exact scar eigenstates $\ket{\mathcal S_m}$ (red lines) and their energies invariant, even across ground-state phase transitions.
 }
 \label{fig:construction}
\end{figure*}

In this Letter, we construct exact scar towers from local constraints induced by Gutzwiller projection~\cite{Gutzwiller1963,Gros1989}; see Fig.~\ref{fig:construction}.
Projecting the ground state of $n$-flavor flat-band Kitaev chains~\cite{Kitaev2001} yields SO$(n)$ symmetry-protected topological (SPT) roots~\cite{Gu2009,Xie2012,Tu2008,Orus2011,Tu2011,Cai2025,Liu2010,Liu2012}, from which a staggered flavor flip generates an equidistant tower. The root and its first descendant exhaust the two-site configurations that are visited by the tower, leaving an orthogonal bond subspace unvisited. 
Deformations confined to this subspace annihilate every tower state bond by bond, preserving the tower regardless of changes in the surrounding spectrum.

We find that the exact $SO(n)$ scar tower can remain unchanged despite changes in ground-state symmetry, order, and the low-energy gap. We demonstrate this in the $SO(4)$ chain, where a simple deformation in the unvisited subspace breaks $SO(4)$ symmetry and drives an exact first-order transition between the SPT and ferromagnetic phases. At the transition point, we construct exact excitations with a quadratic dispersion that establish thermodynamic gaplessness, while the tower wave functions and energies remain unchanged. Exact diagonalization confirms that the tower states remain low-entanglement outliers in the energy spectrum.

\textit{Projected root state and kinetic constraints.}
To construct physical states with controllable local support, we start from the ground state $\ket{\phi}$ of $n$-flavor ($n\ge3$)  Kitaev chains, as shown in Fig.~\ref{fig:construction}(a), 
\begin{equation}
 H_{\rm K} = \sum_{j,\alpha=1}^n \left( -t a_{j,\alpha}^{\dagger}a_{j+1,\alpha} + \Delta a_{j,\alpha}a_{j+1,\alpha} + \mathrm{h.c.} \right),
 \label{eq:kitaev}
\end{equation}
where $a^\dagger_{j, \alpha}$ creates a fermion of flavor $\alpha \in \{1,\dots,n\}$ at lattice site $j$. At the flat-band point $\Delta=t>0$, the Hamiltonian decomposes into commuting nearest-neighbor (NN) Majorana dimers, i.e., $H_{\rm K}=\sum_j h_j$, where $h_j = it \sum_{\alpha} \gamma_{j,\alpha}^{B}\gamma_{j+1,\alpha}^{A}$ with $\gamma_{j,\alpha}^{A}=a_{j,\alpha}+a_{j,\alpha}^{\dagger}$ and $\gamma_{j,\alpha}^{B}=-i(a_{j,\alpha}-a_{j,\alpha}^{\dagger})$. Each NN bond $(j,j+1)$ is locally constrained to the ground-state manifold of $h_j$: the even-parity state $\ket{e_\alpha}=(1+a_{j,\alpha}^\dagger a_{j+1,\alpha}^\dagger)\ket{0}$ and the odd-parity state $\ket{o_\alpha}=(a_{j,\alpha}^\dagger + a_{j+1,\alpha}^\dagger)\ket{0}$. The ground-state wave function of each flavor on bond $(j,j+1)$ therefore lies entirely within the subspace spanned by $\{\ket{e_\alpha},\ket{o_\alpha}\}$.

As a free-fermion state, $\ket{\phi}$ is governed by an extensive set of conserved quantities and does not by itself host quantum many-body scars. Nevertheless, the Gutzwiller projection $\PG$, which imposes a single-occupancy constraint at each site, bridges this free-fermion seed to constrained many-body dynamics. Consequently, the $2^n$-dimensional Fock space at site $j$ is mapped onto the $n$-dimensional physical subspace spanned by $\ket{\alpha_j} = a_{j,\alpha}^\dagger\ket{0}$. Subject to the strict local selection rules imposed by $\PG$, the $n$-flavor unentangled degrees of freedom are transformed into an intricately correlated root state $\ket{\Psi}=\PG\ket{\phi}$; see Fig.~\ref{fig:construction}(a,b).  Specifically, the local pairing structure of $\ket{\phi}$ dictates that, after Gutzwiller projection, only two classes of two-site configurations survive on any nearest-neighbor bond $(j,j+1)$: (i) an SO$(n)$-singlet formed when all flavors are in their even-parity states $\ket{e_\alpha}$, summing to $\sum_{\alpha}\ket{\alpha_j,\alpha_{j+1}}$; and (ii) an antisymmetric state formed when two distinct flavors $\alpha < \beta$ are in their odd-parity states $\ket{o_\alpha}$ and $\ket{o_\beta}$, yielding $\ket{\alpha_j,\beta_{j+1}}-\ket{\beta_j,\alpha_{j+1}}$ due to fermionic anticommutation. Thus, every conditional bond wave function of the projected root $\ket{\Psi}$ is restricted to the singlet and antisymmetric sectors, leaving the symmetric-traceless sector entirely unpopulated, as listed in Table~\ref{tab:bond-states}.

Because the root state $\ket{\Psi}$ leaves the symmetric-traceless sector unpopulated, collective excitations can be naturally generated by projecting into this empty channel. By SO$(n)$ symmetry, any pair of flavors $(\alpha,\beta)$, or their superpositions under SO$(n)$ rotations, can be used to construct an independent scar tower. Without loss of generality, we fix the flavor pair $(\alpha,\beta)=(1,2)$ and define the local flavor-flip operator $q_j = \ket{\alpha^+_j}\bra{\alpha^-_j}$, where $\ket{\alpha^\pm_j} = (\ket{1_j} \pm i\ket{2_j})/\sqrt{2}$. On any bond $(j,j+1)$, the singlet state $\sum_\gamma \ket{\gamma_j,\gamma_{j+1}}$ is symmetric under site exchange, so flipping site $j$ or site $j+1$ yields the same state, causing the staggered combination $(q_j - q_{j+1})$ to cancel identically. Conversely, acting with $(q_j - q_{j+1})$ on an antisymmetric root bond $\ket{\alpha_j,\beta_{j+1}}-\ket{\beta_j,\alpha_{j+1}}$ reverses the exchange sign and maps the state exclusively into the symmetric-traceless sector. 
This motivates the global staggered ladder operator
\begin{equation}
  Q^\dagger = \sum_j (-1)^j q_j,\quad q_j= \ket{\alpha^+_j}\bra{\alpha^-_j},
  \label{eq:Q_operator}
\end{equation} 
which shifts the total momentum by $\pi$ modulo $2\pi$ per application. Since $Q^\dagger$ maps the root state exclusively into the symmetric-traceless sector, the generated excitation is orthogonal to the root, naturally forming the first descendant of the tower.

The root state $\ket{\Psi}$ imposes strict constraints on the action of $Q^\dagger$. First, $q_j^2=0$ prevents two flavor-flips on the same site. Second, acting with $Q^\dagger$ on adjacent sites annihilates the root, $q_j q_{j+1} \ket{\Psi} = 0$, because the combined action requires a pair state of $\ket{\alpha^-_j, \alpha^-_{j+1}}$, which is absent from $\ket{\Psi}$. Since all $q_j$ commute, these conditions restrict any descendant state $|\mathcal{S}_m\rangle\equiv(Q^\dagger)^m \ket{\Psi}$ ($m=1,2,3,..$) to a local bond space spanned only by the root and the first descendant. As summarized in Table~\ref{tab:bond-states},
the tower visits the singlet and antisymmetric
root channels together with only $n-1$
symmetric-traceless bond configurations.
The orthogonal complement has dimension
$n(n-1)/2$ and remains unvisited throughout
the tower, allowing arbitrary local deformations supported on it to leave every
tower state invariant.

\begin{table}[t]
 \centering
 \caption{Two-site configurations on bond $(j,j+1)$ for SO$(n)$ spins. The choice of the orthogonal basis $\ket{\alpha^{\pm}_j}=(\ket{1_j}\pm i\ket{2_j})/\sqrt2$ is arbitrary and does not affect physical observables. $Q^\dagger$ in Eq.~\eqref{eq:Q_operator} denotes the staggered ladder operator generating the tower. 
 }

 \label{tab:bond-states}
 \small
 \setlength{\tabcolsep}{3pt}
 \renewcommand{\arraystretch}{1.2}
 \begin{tabular*}{\linewidth}{@{\extracolsep{\fill}}lcc@{}}
  \toprule
  \toprule
  Bond states on $(j,j+1)$ & Origin & Dimension \\
  \midrule

  $\displaystyle
   \frac{1}{\sqrt n}\sum_{\alpha=1}^n
   \ket{\alpha_j,\alpha_{j+1}}$
  & Root, SO$(n)$-singlet & $1$ \\
  \midrule

  $\displaystyle
   \frac{\ket{\alpha_j,\beta_{j+1}}-\ket{\beta_j,\alpha_{j+1}}}{\sqrt2}$,
  & Root, antisymmetric & $\frac{n(n-1)}{2}$ \\
   ($\alpha<\beta$) & & \\
  \midrule

  \begin{tabular}[c]{@{}l@{}}
   $\ket{\alpha^{+}_j,\alpha^{+}_{j+1}}$ \\[2pt]
   $\frac{\ket{\alpha^{+}_j,r_{j+1}}+\ket{r_j,\alpha^{+}_{j+1}}}{\sqrt2}$ \\
    $(r=3,\ldots,n)$
  \end{tabular}
  &
  \begin{tabular}[c]{@{}c@{}}
  $Q^\dagger$-tower\\
  symmetric-traceless
  \end{tabular}
  & $n-1$ \\
  \midrule

  \begin{tabular}[c]{@{}l@{}}
   States orthogonal to \\
   all states listed above
  \end{tabular}
  & Unvisited
  & $\displaystyle\frac{n(n-1)}{2}$ \\

  \bottomrule
  \bottomrule
 \end{tabular*}
\end{table}

\textit{Local Hamiltonians and scar-preserving deformations.}
The exact scar tower constructed above requires a parent Hamiltonian $H_0$ that hosts these states as an equidistant set of energy eigenstates.
Here we begin with the $L$-site SO$(n)$ AKLT chain~\cite{Affleck1987,Tu2008,Orus2011,Tu2011}:
\begin{equation*}
 H_0
 =J\sum_{j=1}^{L}
 \left[
 \sum_{\alpha<\beta}
 L_j^{\alpha\beta}L_{j+1}^{\alpha\beta}
 +\frac{1}{n}
 \left(
 \sum_{\alpha<\beta}
 L_j^{\alpha\beta}L_{j+1}^{\alpha\beta}
 \right)^2
 \right].
\end{equation*}
where $J>0$ and $L_j^{\alpha\beta}$ ($1\le\alpha<\beta\le n$) are the $n(n-1)/2$ generators of SO$(n)$ with $n \times n$ matrix forms $L^{\alpha\beta} = -i (|\alpha\rangle\langle\beta|-|\beta\rangle\langle \alpha|)$. 
The Hamiltonian can be rewritten as $H_0 = E_0 + 2J \sum_{j=1}^L P_{j,j+1}^{\mathrm{ST}},$ where $E_0 = -J(n-1)L/n$, and $P_{j,j+1}^{\mathrm{ST}}$ denotes the local projector onto the symmetric-traceless sector on bond $(j,j+1)$; 
see Supplemental
Material~\cite{SupplementalMaterial}
for details.
By construction, $P_{j,j+1}^{\mathrm{ST}}$ annihilates all bond states in the singlet and antisymmetric sectors, so the root state $\ket{\Psi}$ is a frustration-free ground state of $H_0$ with energy $E_0$.

By the local constraints established above, the action of $(Q^\dagger)^m$ on any bond $(j,j+1)$ populates only the root or single-flip symmetric-traceless configurations. Since $P_{j,j+1}^{\mathrm{ST}}$ annihilates the root and acts as the identity on these single-flip excitations (each with local energy $2J$), the total energy is simply the sum of local excitations. Summing over bonds, each flipped site contributes $2J$ twice because it is shared by two neighboring bonds. Consequently, every nonzero state $\ket{\mathcal S_m}$ obeys
\begin{equation}
 H_0\ket{\mathcal S_m} = (E_0+4Jm)\ket{\mathcal S_m}.
 \label{eq:tower-energy}
\end{equation}
This directly realizes an exact scar tower with spacing $4J$ and a restricted spectrum-generating algebra~\cite{Mark2020,Moudgalya2020_eta} $[H_0,Q^\dagger]\ket{\mathcal S_m}=4JQ^\dagger\ket{\mathcal S_m}$~\cite{SupplementalMaterial}.

\begin{figure*}[!t]
 \centering
 \includegraphics[width=\textwidth]{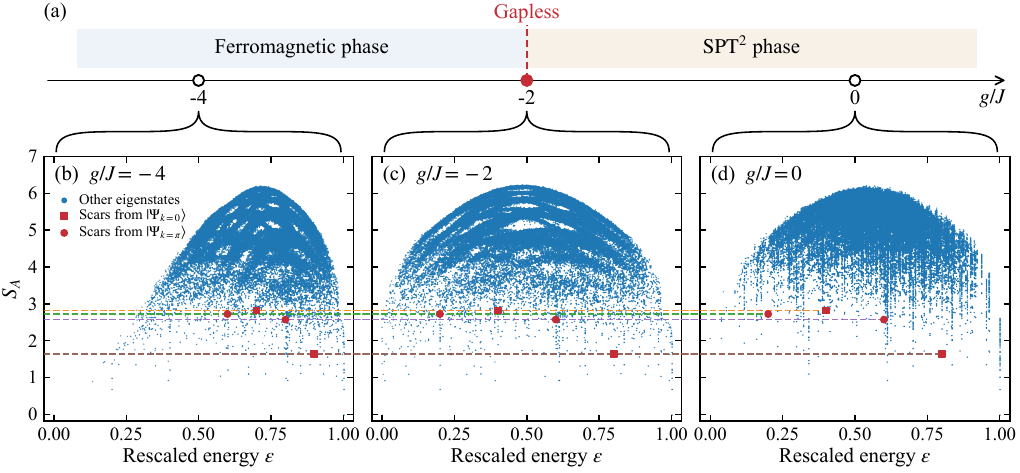}
 \caption{
 Ground-state phase diagram and invariant scar states in the SO$(4)$ AKLT chain under a polarization-selective deformation; see Eq.~\eqref{eq:Hg}.
 (a) The ferromagnetic phase ($g<-2J$) and SPT$^{2}$ phase ($g>-2J$) are separated by a gapless first-order transition at $g=-2J$.
 (b)--(d) Half-chain entanglement entropy $S_A$ versus rescaled energy $\varepsilon=[E-E_{\min}(g)]/[E_{\max}(g)-E_{\min}(g)]$ for all $k=0$ eigenstates at $L=10$ and $g/J=-4,-2,0$.
 Blue dots denote other thermal eigenstates; red squares ($m=2,4$) and circles ($m=1,3$) mark tower states descending from the $k=0$ and $k=\pi$ roots, respectively.
 }
 \label{fig:numerical-spectra}
\end{figure*}

The SO($n$) exact scar tower constructed using Gutzwiller-projected fermions now is transparent. Specifically, the root space is spanned by the projected states $\ket{\Psi}$ that satisfy the local bond constraints, and repeated applications of $Q^\dagger$ generate the tower states $\ket{\mathcal{S}_m} \equiv (Q^\dagger)^m \ket{\Psi}$. Moreover, as pointed out in Refs.~\cite{Tu2008,Cai2025}, the degeneracy of these root states depends on the parity of $n$. For odd $n$ [$SO(2l+1)$], $\PG$ yields a unique root state with total momentum $k=0$. For even $n$ [$SO(2l)$], such as the $SO(4)$ chain studied below, two root states with $k=0$ and $k=\pi$, corresponding to periodic and antiperiodic boundary conditions in Eq.~\eqref{eq:kitaev}~\cite{Cai2025}, survive under the Gutzwiller projection. These two states span the root space and give rise to two intertwined scar towers in $SO(2l)$ chains, as shown in Fig.~\ref{fig:construction}(b).

Crucially, while the projector term $P_{j,j+1}^{\mathrm{ST}}$ penalizes the entire symmetric-traceless sector, the scar tower visits only $n-1$ states within it, leaving an orthogonal bond subspace of dimension $n(n-1)/2$ completely unvisited, as listed in Table~\ref{tab:bond-states}. We can therefore introduce the nearest-neighbor Hamiltonian deformation such as 
$H(g)=H_0+g\sum_{j=1}^L O_{j,j+1}$, where $O_{j,j+1}$ is a Hermitian operator supported solely on this unvisited subspace. Consequently, every tower state is annihilated by such a deformation term bond by bond, i.e., $O_{j,j+1}\ket{\mathcal S_m}=0$.  
Regardless of the specific form of $O_{j,j+1}$ and the coupling $g$, the tower wave functions $\ket{\mathcal S_m}$, absolute energies $E_0+4Jm$, and entanglement entropies remain strictly invariant.

\textit{Exact scar tower against phase transition.---}
While deformations supported on the unvisited bond subspace leave the scar tower invariant, they can fundamentally reshape the low-energy physics and drive ground-state phase transitions. 
We consider a periodic SO$(n)$ AKLT chain under a polarization-selective deformation:
\begin{equation}
 H(g)=H_0+g\sum_{j=1}^L \sigma^{-}_{j}\sigma^{-}_{j+1}, \quad \sigma^{-}_{j}=\ket{\alpha^-_j}\bra{\alpha^-_j}.
 \label{eq:Hg}
\end{equation}
Because the operator $\sigma^{-}_{j}\sigma^{-}_{j+1}$ acts within the symmetric-traceless sector, the energy cost of bond configurations is shifted by $g$ for the polarized states $\ket{\alpha^-_j, \alpha^-_{j+1}}$ relative to other symmetric-traceless states. This allows us to systematically tune the coupling $g$ and drive a ground-state phase transition while leaving the scar tower completely unchanged.

The bond states listed in Table~\ref{tab:bond-states} make the ground-state competition explicit. Measuring bond energies relative to $E_0/L$, the singlet and antisymmetric root channels have energy zero, while the $n-1$ symmetric-traceless channels visited by the tower have energy $2J$. In the unvisited subspace, the deformation lowers or raises the energy of $\ket{\alpha^-_j,\alpha^-_{j+1}}$ to $2J+g$, leaving all other states at $2J$.

For $g>-2J$, all channels outside the root sector have positive bond energies. The projected roots therefore remain the frustration-free ground states with energy $E_0$. For $g<-2J$, the polarized pair $\ket{\alpha^-_j,\alpha^-_{j+1}}$ in the unvisited subspace becomes the lowest-energy bond state. The fully polarized state $\ket F=\bigotimes_j\ket{\alpha^-_j}$ occupies this state on every bond and is an exact eigenstate with energy $E_0+L(2J+g)$. It minimizes every bond simultaneously and is thus the unique ground state in this regime. The ground-state energy has a kink at $g=-2J$, signaling a sharp first-order phase transition between the topological SPT phase and the ferromagnetic phase, as shown in Fig.~\ref{fig:numerical-spectra} (a).

At the transition point $g=-2J$, the projected roots and $\ket F$ are degenerate ground states. Furthermore, we find that this transition point hosts gapless single-flip excitations from the degenerate ground-state manifold. 
Flipping any site $j$ in the polarized background $\ket F$ into an orthogonal flavor $r\in\{3,\dots,n\}$ creates a local defect.
A Fourier superposition of these defects defines $\ket{k,r}=L^{-1/2}\sum_j e^{ikj}\ket{r_j}\bra{\alpha^-_j}\ket F$. 
On the two bonds adjacent to the defect, the effect of deformation vanishes, while $H_0$ gives a total diagonal excitation energy $2J$ and moves the defect to either neighboring site with amplitude $J$. This yields the exact dispersion $\varepsilon(k)=2J(1+\cos k)$; see Supplemental
Material~\cite{SupplementalMaterial}
for details.
Near $k=\pi$, this gives $n-2$ degenerate quadratic gapless branches $\varepsilon(\pi+q) \simeq Jq^2$. Therefore, the excitations at $k = \pi \pm 2\pi/L$ scale as $O(L^{-2})$, providing an upper bound on the finite-size gap of $0 < \Delta_L(-2J) \le \frac{4\pi^2 J}{L^2} + O(L^{-4}).$
As these are exact eigenstates orthogonal to the ground-state manifold, this rigorously proves thermodynamic gaplessness at the transition. Throughout the transition, the scar tower remains invariant in wave functions and energies, standing out as low-entanglement outliers embedded in the reorganizing spectrum.

\textit{Numerical verification.}
We perform exact diagonalization on the $L=10$  SO$(4)$ chain to verify our analytical results. The scar tower states $\ket{\mathcal{S}_m}$ are generated by iteratively applying the ladder operations $Q^\dagger$ on top of the root state $\ket{\Psi}$. 
Fig~\ref{fig:numerical-spectra}(b)--(d) plots the half-chain entanglement entropy $S_A$ against the rescaled energy $\varepsilon\equiv[E-E_{\min}(g)]/[E_{\max}(g)-E_{\min}(g)]$ for all $k=0$ eigenstates at $g/J=-4$, $-2$, and $0$, where $E_{\min}(g)$ and $E_{\max}(g)$ are the lowest and highest eigenvalues of $H(g)$ in Eq.~\eqref{eq:Hg}. Because $n=4$ is even, there are two degenerate root states carrying momenta $k=0$ and $k=\pi$, respectively. The marked $k=0$ tower states with $m=1,3$ ($m=2,4$) descend from $\ket{\Psi_{k=\pi}}$ ($\ket{\Psi_{k=0}}$), shown as circles (squares) in Fig~\ref{fig:numerical-spectra}.

The surrounding spectrum reorganizes qualitatively
across the transition, including changes in the
distribution of states at low rescaled energies.
The four marked scar states remain unaffected:
their absolute energies and
half-chain entanglement entropies are identical
in all three panels.
They remain embedded as low-entanglement outliers
on both sides of the transition and at the
gapless point. For a larger chain with $L=40$, the entanglement
entropy of a representative mid-spectrum scar
state exhibits approximately logarithmic growth
with subsystem size~\cite{SupplementalMaterial}.

\textit{Conclusion.}
Gutzwiller projection of flat-band Kitaev chains naturally induces local bond exclusions in $\mathrm{SO}(n)$ spin chains. These constraints restrict the available two-site configurations, allowing for the construction of equidistant scar towers. Because the entire tower visits only a small subset of bond states, deformations confined to the unvisited subspace preserve the tower wave functions and energies while driving ground-state transitions. We demonstrated an exact first-order transition with a gapless critical point where the scar tower persists. Our results show that local kinetic constraints can isolate high-energy exact scars from the low-energy physics of phase transitions.

For further study, it will be interesting to extend this Gutzwiller-projection mechanism beyond the flat-band limit or to higher-dimensional lattices to yield new scar towers. It is also worth exploring whether these projection-induced constraints can protect coherent dynamics in driven or open systems\cite{Wang2024,Jiang2026}, or sustain scar revivals when the tower is weakly coupled to a thermal background.

\begin{acknowledgments}
We acknowledge the support by the National Natural Science Foundation of China (Grant NSFC-12494594, NSFC-12574150, NSFC-12174428, NSFC-12504180), the Ministry of Science and Technology (Grant No. 2022YFA1403900), the Chinese Academy of Sciences Project for Young Scientists in Basic Research (2022YSBR-048), and the start-up funding from ShanghaiTech University.
The numerical calculations used the ITensor and QuSpin libraries~\cite{Fishman2022,Weinberg2019}.
\end{acknowledgments}

\bibliography{refs}

% ======================================================================
% SUPPLEMENTAL MATERIAL
% ======================================================================

\clearpage
\onecolumngrid
\setcounter{secnumdepth}{2}
\setcounter{section}{0}
\setcounter{equation}{0}
\setcounter{figure}{0}
\renewcommand{\theequation}{S\arabic{equation}}
\renewcommand{\thefigure}{S\arabic{figure}}
\renewcommand{\theHfigure}{S\arabic{figure}}
\renewcommand{\theHequation}{S\arabic{equation}}

\begin{center}
{\large\bfseries Supplemental Material for
``Projecting Fermions into Exact Scar Towers:\\ 
Local Protection across Phase Transitions''}
\end{center}

\vspace{1.cm}

\twocolumngrid

This Supplemental Material provides additional results supporting the main text.
Section~\ref{sec:sm-projector} derives the bond-projector representation
of the SO$(n)$ bilinear-biquadratic Hamiltonian.
Section~\ref{sec:sm-generating} establishes the exact tower energies
and the restricted spectrum-generating algebra, including their
invariance under tower-preserving deformations.
Section~\ref{sec:sm-phase} proves the ground-state transition and
the spectral gap results, using exact single-defect excitations to establish gapless excited states at the transition.
Section~\ref{sec:sm-entanglement} presents the $L=40$ entanglement profile and derives a logarithmic upper bound on tower entanglement.

% ======================================================================
% I. PROJECTOR REPRESENTATION
% ======================================================================

\section{From the SO$(n)$ chain to the bond projector}
\label{sec:sm-projector}

We derive the projector representation of $H_0$ directly in the physical SO$(n)$ vector basis. The generators act as
\begin{equation}
 L^{\alpha\beta}\ket\mu
 =-i\bigl(
 \delta_{\beta\mu}\ket\alpha
 -\delta_{\alpha\mu}\ket\beta
 \bigr),
 \qquad \alpha<\beta.
 \label{eq:sm-generator-action}
\end{equation}
Defining
$X_{j,j+1}=\sum_{\alpha<\beta}
L_j^{\alpha\beta}L_{j+1}^{\alpha\beta}$,
we obtain
\begin{equation}
 X_{j,j+1}\ket{\mu_j,\nu_{j+1}}=
 \begin{cases}
 \ket{\nu_j,\mu_{j+1}},&\mu\ne\nu,\\[2pt]
 \ket{\mu_j,\mu_{j+1}}
 -\displaystyle\sum_{\gamma=1}^{n}
 \ket{\gamma_j,\gamma_{j+1}},&\mu=\nu.
 \end{cases}
 \label{eq:sm-X-action}
\end{equation}
The singlet $\sum_\mu\ket{\mu_j,\mu_{j+1}}$ therefore has
eigenvalue $1-n$. Antisymmetric states have eigenvalue $-1$,
whereas symmetric states orthogonal to the singlet have
eigenvalue $1$. The bond Hamiltonian
$JX_{j,j+1}+(J/n)X_{j,j+1}^2$ consequently has energy
$-J(n-1)/n$ in the singlet and antisymmetric sectors, and
$J(n+1)/n$ in the symmetric-traceless sector. Thus,
\begin{equation}
 JX_{j,j+1}+\frac{J}{n}X_{j,j+1}^2
 =-\frac{J(n-1)}{n}\Id
 +2JP_{j,j+1}^{\mathrm{ST}},
 \label{eq:sm-bond-parent}
\end{equation}
where $P_{j,j+1}^{\mathrm{ST}}$ projects onto the
symmetric-traceless sector. Summing over the $L$ bonds gives
\begin{equation}
H_0=E_0+2J\sum_jP_{j,j+1}^{\mathrm{ST}},
 \qquad E_0=-\frac{J(n-1)L}{n}.
 \label{eq:sm-AKLT}
\end{equation}
For $J>0$, any state annihilated by every
$P_{j,j+1}^{\mathrm{ST}}$ minimizes each bond term and is
a frustration-free ground state with energy $E_0$.

% ======================================================================
% II. EXACT LADDER
% ======================================================================

\section{Exact ladder and restricted spectrum-generating algebra}
\label{sec:sm-generating}

We derive the exact tower energies and the restricted
spectrum-generating algebra stated in the main text.
The scar tower is generated by a staggered ladder operator defined in the main text as 
$$Q^\dagger = \sum_j (-1)^j q_j,\quad q_j= \ket{\alpha^+_j}\bra{\alpha^-_j}.$$
The projected root $\ket{\Psi}$ contains only singlet
and antisymmetric bond components, which implies
\begin{align}
 P_{j,j+1}^{\mathrm{ST}}\ket{\Psi}&=0,\nonumber\\
 q_jq_{j+1}\ket{\Psi}&=0,\nonumber\\
 P_{j,j+1}^{\mathrm{ST}}(q_j-q_{j+1})\ket{\Psi}
 &=(q_j-q_{j+1})\ket{\Psi}.
 \label{eq:sm-local-input}
\end{align}
Here the staggered flip annihilates the singlet
and maps antisymmetric bond states into the
symmetric-traceless sector.

Since $q_j^2=0$ and operators on different sites
commute, the generating state is
\begin{equation}
 \begin{aligned}
 \ket{\Psi(z)}
 &=e^{zQ^\dagger}\ket{\Psi}=\sum_{m=0}^{L}\frac{z^m}{m!}\ket{\mathcal S_m},
 \end{aligned}
 \label{eq:sm-generator}
\end{equation}
where, as defined in the main text, $\ket{\mathcal S_m}=(Q^\dagger)^m|\Psi\rangle.$
On each bond, Eq.~\eqref{eq:sm-local-input}
eliminates the quadratic term and makes
$P_{j,j+1}^{\mathrm{ST}}$ retain only the
single-flip contribution.
Summing over bonds counts each flipped site twice,
once from each adjacent bond, giving
\begin{equation}
 \sum_jP_{j,j+1}^{\mathrm{ST}}\ket{\Psi(z)}
 =2z\partial_z\ket{\Psi(z)}.
 \label{eq:sm-global-counting}
\end{equation}
Using $H_0=E_0+2J\sum_jP_{j,j+1}^{\mathrm{ST}}$
and comparing coefficients of equal powers of $z$, we obtain
\begin{equation}
 H_0\ket{\mathcal S_m}
 =(E_0+4Jm)\ket{\mathcal S_m}
 \label{eq:sm-tower-energy}
\end{equation}
for every nonzero tower state.
Since $Q^\dagger\ket{\mathcal S_m}
=\ket{\mathcal S_{m+1}}$, this immediately yields
\begin{equation}
 [H_0,Q^\dagger]\ket{\mathcal S_m}
 =4JQ^\dagger\ket{\mathcal S_m},
 \label{eq:sm-restricted-sga}
\end{equation}
which realizes a restricted spectrum-generating algebra.

For the tower-preserving deformations,
$O_{j,j+1}\ket{\mathcal S_m}=0$ for every $j$
and $m$.
Consequently, both the tower energies and the
restricted spectrum-generating algebra remain unchanged
for $H(g)=H_0+g\sum_jO_{j,j+1}$.

% ======================================================================
% III. GROUND STATES AND GAPS
% ======================================================================

\section{Ground-state phase transition and spectral gaps of deformation}
\label{sec:sm-phase}

We consider the deformed Hamiltonian
$H(g)=H_0+g\sum_{j=1}^{L}\sigma^-_j\sigma^-_{j+1}$
on an even periodic chain of length $L$, and
$\sigma^-_j=\ket{\alpha^-_j}\bra{\alpha^-_j}$.
We use $H_0=E_0+2J\sum_jP_{j,j+1}^{\mathrm{ST}}$,
with $E_0=-J(n-1)L/n$.
The gap $\Delta_L(g)$ denotes the smallest strictly positive
excitation energy above the full ground-state space.

\subsection{Bond energies and ground-state transition}

The ferromagnetic bond state
$\ket{\alpha^-_j,\alpha^-_{j+1}}$ is symmetric and
orthogonal to the SO$(n)$ singlet, so its projector
$\sigma^-_j\sigma^-_{j+1}$ is a subprojector of
$P_{j,j+1}^{\mathrm{ST}}$.
The shifted bond Hamiltonian therefore decomposes into
two orthogonal projectors:
\begin{equation}
 \begin{aligned}
 2JP_{j,j+1}^{\mathrm{ST}}+g\sigma^-_j\sigma^-_{j+1}=
 &2J\left(P_{j,j+1}^{\mathrm{ST}}
 -\sigma^-_j\sigma^-_{j+1}\right)\\
 &+(2J+g)\sigma^-_j\sigma^-_{j+1}.
 \end{aligned}
 \label{eq:sm-phase-bond}
\end{equation}
The two projectors on the right-hand side act differently
on the competing bond states. Both annihilate the singlet
and antisymmetric components of the projected roots,
giving zero shifted bond energy. On the ferromagnetic bond state,
\begin{align}
 \left(P_{j,j+1}^{\mathrm{ST}}
 -\sigma^-_j\sigma^-_{j+1}\right)
 \ket{\alpha^-_j,\alpha^-_{j+1}}&=0,\nonumber\\
 \sigma^-_j\sigma^-_{j+1}
 \ket{\alpha^-_j,\alpha^-_{j+1}}
 &=\ket{\alpha^-_j,\alpha^-_{j+1}},
\end{align}
so its shifted bond energy is $2J+g$.
All other symmetric-traceless states orthogonal to this
bond state have shifted bond energy $2J$.

For $g>-2J$, both coefficients in
Eq.~\eqref{eq:sm-phase-bond} are positive, so the
ground-state space remains the same as at $g=0$,
with energy $E_0$.
The system also remains gapped in the thermodynamic
limit for every fixed $g>-2J$.
For $g<-2J$, the unique lowest-energy bond state is
$\ket{\alpha^-_j,\alpha^-_{j+1}}$.
Occupying it on every bond gives the unique ground state
$\ket F=\bigotimes_j\ket{\alpha^-_j}$, with energy
$E_0+L(2J+g)$.

Thus, the ground-state energy density is
$E_{GS}/L=E_0/L+\min\{0,2J+g\}$.
Its derivative with respect to $g$ jumps from $1$ to $0$
at $g=-2J$, establishing a first-order ground-state transition.

\subsection{Gapless transition at the transition point}

For $g<-2J$, subtracting the ferromagnetic ground-state energy leads to 
$H(g)-E_0-L(2J+g)=
 2J\sum_j\left(P_{j,j+1}^{\mathrm{ST}}
 -\sigma^-_j\sigma^-_{j+1}\right)
 -(2J+g)\sum_j\left(\Id-\sigma^-_j\sigma^-_{j+1}\right)
 \ge0.$
Both sums have nonnegative coefficients and
nonnegative terms.
The second sum counts nonpolarized bonds. It vanishes on $\ket F$,
while every other product configuration contains at least two such bonds, giving
$\sum_j(\Id-\sigma^-_j\sigma^-_{j+1})
\ge2(\Id-\ket F\bra F)$.
Together with the nonnegative first sum, this implies $\Delta_L(g)\ge-2(2J+g)>0$ for $g<-2J$.

We next construct exact single-defect excitations above the polarized state $\ket F$.
In order to do this, we flip one
$\ket{\alpha^-_j}$ at site $j$ in the polarized background $\ket F$ with $\ket{r_j}$, $r=3,\ldots,n$. Their normalized
momentum superpositions are
\begin{equation}
 \ket{k,r}
 =\frac{1}{\sqrt L}\sum_{j=1}^{L}
 e^{ikj}\ket{r_j}\bra{\alpha^-_j}\ket F,
 \qquad k=\frac{2\pi\ell}{L}.
 \label{eq:sm-phase-defect-momentum}
\end{equation}
At $g=-2J$, the polarized state $\ket F$ is a ground
state with energy $E_0$. 
A defect of flavor $r=3,\ldots,n$
changes only the two bonds adjacent to its site.
On either bond, the defect configuration is orthogonal
to the singlet, so
\begin{equation}
 P_{j,j+1}^{\mathrm{ST}}\ket{r_j,\alpha^-_{j+1}}
 =\frac12\left(
 \ket{r_j,\alpha^-_{j+1}}
 +\ket{\alpha^-_j,r_{j+1}}
 \right).
\end{equation}
Meanwhile, $\sigma^-_j\sigma^-_{j+1}$ annihilates it.
Each adjacent bond therefore contributes a diagonal
amplitude $J$ and a hopping amplitude $J$. All other
bonds contribute zero to $H(-2J)-E_0$. Consequently,
\begin{equation}
 \begin{aligned}
 &[H(-2J)-E_0]\ket{r_j}\bra{\alpha^-_j}\ket F\\
 &\quad=2J\ket{r_j}\bra{\alpha^-_j}\ket F\\
 &\qquad+J\ket{r_{j-1}}\bra{\alpha^-_{j-1}}\ket F
       +J\ket{r_{j+1}}\bra{\alpha^-_{j+1}}\ket F.
 \end{aligned}
 \label{eq:sm-phase-critical-hop}
\end{equation}
The single-defect subspace is thus invariant.
Fourier transformation gives the exact eigenvalue equation
\begin{equation}
 [H(-2J)-E_0]\ket{k,r}
 =2J(1+\cos k)\ket{k,r}.
 \label{eq:sm-phase-critical-dispersion}
\end{equation}

Therefore, the single-mode excited states form $n-2$ exact excitation branches, with
$\varepsilon(\pi+q)=Jq^2+O(q^4)$ near $k=\pi$.
For even $L$, the states at $k=\pi$ belong to the
ground-state space. To establish gaplessness, we instead
take $k=\pi\pm2\pi/L$. These states have strictly positive
energies and are therefore orthogonal to the entire
ground-state space, giving
\begin{equation}
 \begin{aligned}
 0<\Delta_L(-2J)
 &\le 2J\left[1-\cos\left(\frac{2\pi}{L}\right)\right]\\
 &=\frac{4\pi^2J}{L^2}+O(L^{-4}).
 \end{aligned}
 \label{eq:sm-phase-critical-gap}
\end{equation}
The upper bound vanishes as $L\to\infty$, proving
thermodynamic gaplessness for every fixed $n\ge3$.
Note that the single-defect subspace is exactly invariant under the full Hamiltonian. The resulting dispersion therefore receives no corrections from coupling to the degenerate SPT ground states.

% ======================================================================
% V. ENTANGLEMENT
% ======================================================================

\section{Logarithmic entanglement growth}
\label{sec:sm-entanglement}

To examine the entanglement of scar states in a
larger system, we extend the calculation to a
periodic $SO(4)$ chain with $L=40$.
We consider a representative scar state in the
middle of the energy spectrum and calculate its
entanglement entropy $S_A$ for a subsystem of
$l_A$ consecutive physical sites.
For comparison, we also show the entropy of
an MPS ground state of $H_0$.

\begin{figure}[ht]
 \centering
 \includegraphics[width=0.5\textwidth]
 {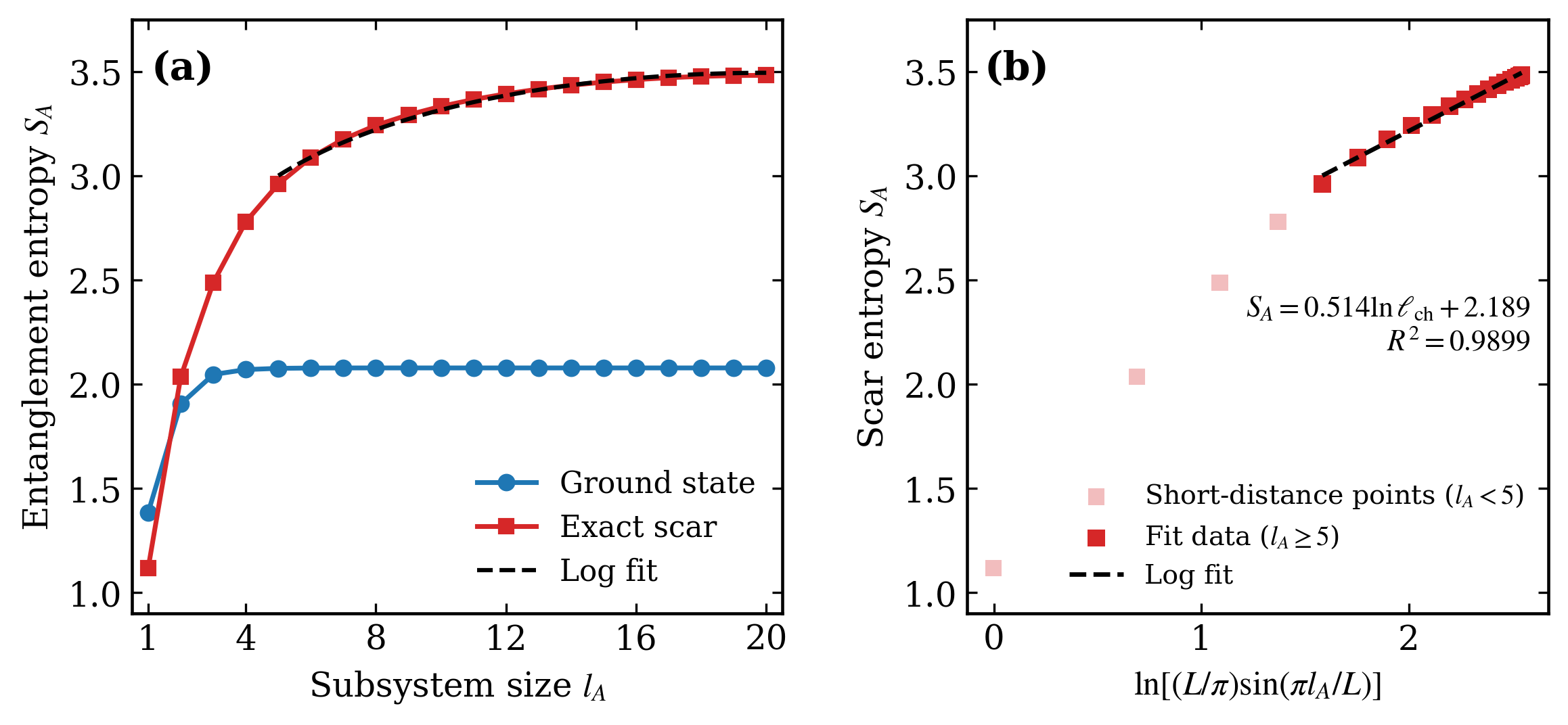}
 \caption{
 Logarithmic entanglement growth of a
 mid-spectrum scar state in the periodic
 $L=40$ $SO(4)$ chain.
 (a) Entanglement entropy versus the number
 $l_A$ of consecutive sites in the subsystem.
 The MPS ground-state entropy rapidly
 saturates, while the scar-state entropy
 increases slowly.
 The dashed curve is the fit to
 Eq.~\eqref{eq:sm-log-fit}.
 (b) The same scar-state data plotted
 against the logarithm of the chord length.
 The dashed line fits the data for
 $5\le l_A\le L/2$, yielding
 $a=0.514$ and $b=2.189$.
 The four lighter points with $l_A<5$
 are excluded from the fit.
 }
 \label{fig:sm-entanglement}
\end{figure}

As shown in Fig.~\ref{fig:sm-entanglement}(a),
the ground-state entropy rapidly saturates
as the subsystem size increases.
The mid-spectrum scar state instead exhibits
a slow increase in entanglement.
To quantify this dependence while accounting
for the periodic finite-size geometry,
we fit the data for $5\le l_A\le L/2$ to
\begin{equation}
 S_A^{\rm fit}(l_A)
 =
 a\ln\!\left[
 \frac{L}{\pi}
 \sin\!\left(\frac{\pi l_A}{L}\right)
 \right]+b.
 \label{eq:sm-log-fit}
\end{equation}
The fit gives $a=0.514$ and $b=2.189$,
with $R^2=0.9899$ and a root-mean-square
residual of $0.0153$.
Figure~\ref{fig:sm-entanglement}(b) displays
the same scar-state data against the logarithm
of the chord length, revealing an approximately
linear dependence over the fitting interval.
These results support logarithmic growth
of the mid-spectrum scar state's entanglement
with subsystem size at $L=40$.
The chord-length expression is used as
a finite-size fitting form.

This numerical behavior is consistent with
a general entanglement bound for the tower.
For the bipartition into $A$ and $B$,
write $Q^\dagger=Q_A^\dagger+Q_B^\dagger$.
Since the two operators act on different
regions, they commute, giving
\begin{equation}
 \ket{\mathcal S_m}
 =
 \sum_{p=0}^{m}\binom{m}{p}
 (Q_A^\dagger)^p
 (Q_B^\dagger)^{m-p}\ket{\Psi}.
 \label{eq:sm-Schmidt-expansion}
\end{equation}
Let $r_{\rm root}$ be the Schmidt rank
of the root across this bipartition.
Each term is obtained by separate operations
on $A$ and $B$, so its Schmidt rank cannot
exceed $r_{\rm root}$.
Moreover, $q_j^2=0$ implies
$(Q_A^\dagger)^p=0$ for $p>l_A$.
There are therefore at most
$\min\{m,l_A\}+1$ nonzero terms.
The entropy of the normalized, nonzero
tower state consequently satisfies
\begin{equation}
 S_A^{(m)}
 \le
 \ln r_{\rm root}
 +\ln\!\left[\min\{m,l_A\}+1\right].
 \label{eq:sm-entanglement-bound}
\end{equation}
For a contiguous subsystem, the MPS root
has bounded Schmidt rank at fixed $n$,
independently of $L$ and $l_A$.
The tower entropy is thus at most logarithmic
in subsystem size, consistent with the
numerical profile above.
Both the bound and the entropy profile
remain unchanged under tower-preserving
deformations, because the tower wave
functions are unchanged.

\end{document}